\documentclass[11pt]{article}
\begin{document}
\begin{flushright}
SJSU/TP-99-20\\
June 1999\end{flushright}
\vspace{1.7in}
\begin{center}\Large{\bf Quantum Probability from Decision Theory?}\\
\vspace{1cm}
\normalsize\ J. Finkelstein\footnote[1]{
        Participating Guest, Lawrence Berkeley National Laboratory\\
        \hspace*{\parindent}\hspace*{1em}
        e-mail: JLFINKELSTEIN@lbl.gov}\\
        Department of Physics\\
        San Jos\'{e} State University\\San Jos\'{e}, CA 95192, U.S.A
\end{center}
\begin{abstract}
Deutsch has recently (in quant-ph/9906015) offered a justification,
based only on the non-probabilistic axioms of quantum theory and of 
classical decision theory, for the use of the standard quantum probability
rules. In this note, this justification is examined.
\end{abstract}
\newpage
The origin, and indeed the meaning, of the probability rules of quantum
theory are often considered mysterious.  In a recent article
(ref. \cite{D}, hereinafter denote by [D]), David Deutsch has offered a
proof, based only on the non-probabilistic axioms of quantum theory
and of classical decision theory, that a ``rational decision maker''
would act as if the standard probability rules of quantum theory were 
true.  (See also ref. \cite{P}.)
This result could then be taken to justify these standard rules,
without the need to introduce any probabilistic axiom, or even the need to
have any definition of probability.  It is the purpose of this note to
examine the proof given in [D].

[D] considers a decision maker who is offered the following game:
an observable $\hat{X}$ is to be measured on a quantum system which is in the
state $|\Psi \rangle $; the utility of the payoff offered is numerically
equal to the result of the measurement.  The value of this game is denoted
by ${\cal V}[|\Psi \rangle ]$, and the proof given in [D] is that,
even without any probabilistic axioms, it follows that
\begin{equation}
   {\cal V}[|\Psi \rangle ] = \langle \Psi |\hat{X}|\Psi \rangle.  
   \label{a}\end{equation}
The RHS of Eq.~\ref{a} is of course the value of the game that would 
follow from the standard quantum probability rules; the derivation of 
Eq.~\ref{a} without invoking any probability axioms can thus be considered
a justification of those rules.

I wish to examine a central part of the proof offered in [D], in which
the state $|\Psi\rangle $ is an equal-amplitude superposition of two 
eigenstates of $\hat{X}$. I will use the notation employed in [D],
that $x$ is an eigenvalue, and $|x\rangle $ the corresponding eigenstate,
of the operator $\hat{X}$.  Then the states we will be concerned with
can be written
\begin{equation}
    |\Psi\rangle =\mbox{\small $\frac{1}{\sqrt{2}}$}(|x_{1}\rangle +
    |x_{2}\rangle); \label{b}\end{equation}
this is also Eq.\ 7 of [D].  The non-probabilistic part of quantum theory
implies that the outcome of the measurement of $\hat{X}$ on the state
given in Eq.~\ref{b} is surely either $x_1$ or $x_2$ (but of course does
not specify which one); from this fact, together with axioms of classical
decision theory, [D] deduces that (Eq.~11 in [D])
\begin{equation}
     {\cal V}[\mbox{\small $\frac{1}{\sqrt{2}}$}(|x_{1}\rangle +
     |x_{2}\rangle)] =
     \mbox{\small $\frac{1}{2}$}(x_{1}+x_{2}). \label{c}\end{equation}
This result (our Eq.~\ref{c}) is called the ``pivotal result'' of [D].

But how could this result follow from the argument made in [D]?
Let $r$ denote the outcome of the measurement of $\hat{X}$ on the state
given in Eq.~\ref{b}; then the only requirement of quantum theory used in 
the proof of Eq.~\ref{c} is that either $r=x_1$ or $r=x_2$.  Now suppose
(consistent with that requirement) that the value of $r$ were in fact
determinate, and were given (in the case $|x_{1}|\neq |x_{2}|$) by
\begin{eqnarray}
    r & = & x_{1} \;\;\;\mbox{ if } |x_{1}| > |x_{2}| \nonumber \\
    r & = & x_{2} \;\;\;\mbox{ if } |x_{2}| > |x_{1}|.\label{a1}\end{eqnarray}
When the outcome of the measurement is determinate, the value of the 
game is simply equal to the outcome, so if, for example, we set
$x_{1}=1$ and $x_{2}=0$, this supposition gives $r=1$, and so the LHS of 
Eq.~\ref{c} would be 1
while the RHS of that equation would be 0.5.

The proof of Eq.~\ref{c} (Eq.\ 11 of [D]) begins with Eq.\ 8 of [D].
That equation, specialized to the case of an equal-amplitude 
superposition of two eigenstates, reads
\begin{equation}
   {\cal V}[\mbox{\small $\frac{1}{\sqrt{2}}$}(|x_{1}+k\rangle +
   |x_{2}+k\rangle)] =
   k+{\cal V}[\mbox{\small $\frac{1}{\sqrt{2}}$}(|x_{1}\rangle +
   |x_{2}\rangle)], \label{d8}\end{equation}
and is supposed to be true for any $k$.  However, if the notation in
Eq.~\ref{d8} is taken to mean the same as that in Eqs.~\ref{a} and \ref{b}
(e.g., that $|x_{1}+k\rangle$ is the eigenstate corresponding to eigenvalue
$x_{1}+k$),
the game on the LHS of Eq.~\ref{d8} may not even be possible.  In writing
Eq.~\ref{b}, we have assumed that the operator $\hat{X}$ has eigenvalues
$x_1$ and $x_2$, but since nothing else was assumed about the spectrum of
this operator, it might not have eigenvalues $x_{1}+k$ and $x_{2}+k$,
and so the state required by the game on the LHS of Eq.~\ref{d8} might
not exist.  Furthermore, in the case in which this state {\em does} happen to
exist, the argument given in [D] would allow this equation to be incorrect.
Consider the example given in Eq.~\ref{a1}, and take $x_{1}=1$,
$x_2=0$, and $k=-1$.  Then since, by Eq.~\ref{a1}, 
${\cal V}[\frac{1}{\sqrt{2}}(|0\rangle +|-1\rangle)]=-1$ and
${\cal V}[\frac{1}{\sqrt{2}}(|1\rangle +|0\rangle)]=+1$,
Eq.~\ref{d8} reads $-1=0$.

On the other hand, it is possible to re-interpret the notation used in
Eq.~\ref{d8} in a way that makes it correct.  Let me set
$|x_{1},x_{2}\rangle =:\frac{1}{\sqrt{2}}(|x_{1}\rangle +|x_{2}\rangle)$,
and define
   \[{\cal V}[|x_{1},x_{2}\rangle;\;x_{1}\rightarrow u{_1},\;
   x_{2}\rightarrow u_{2}]\]
to be the value of the following game: $\hat{X}$ is measured on the
state $|x_{1},x_{2}\rangle$; if the outcome is $x_1$, the utility is 
$u_1$, while if the outcome is $x_2$, the utility is $u_2$. The relation
between this notation and the original notation is
\begin{equation}
  {\cal V}[\mbox{\small $\frac{1}{\sqrt{2}}$}(|x_{1}\rangle +
  |x_{2}\rangle)] = {\cal V}[|x_{1},x_{2}\rangle;\;x_{1}\rightarrow x{_1},\;
   x_{2}\rightarrow x_{2}]. \label{e}\end{equation}
Now if we interpret the LHS of Eq.~\ref{d8} not as
${\cal V}[|x_{1}+k,x_{2}+k\rangle;\;x_{1}+k\rightarrow x{_1}+k,\;
   x_{2}+k\rightarrow x_{2}+k]$, but rather as
${\cal V}[|x_{1},x_{2}\rangle;\;x_{1}\rightarrow x{_1}+k,\;
   x_{2}\rightarrow x_{2}+k]$, we can re-write Eq.~\ref{d8} as
\begin{equation}
  {\cal V}[|x_{1},x_{2}\rangle;\;x_{1}\rightarrow x{_1}+k,\;
   x_{2}\rightarrow x_{2}+k] = k+{\cal V}[|x_{1},x_{2}\rangle;\;x_{1}
   \rightarrow x{_1},\;x_{2}\rightarrow x_{2}]. \label{d8c}\end{equation}
This equation is meaningful (because the same state $|x_{1},x_{2}\rangle$
is required in the two games), and moreover it is correct.  So perhaps
this is the way we were intended to understand Eq.\ 8 of [D].

We can now attempt to complete the derivation of Eq.~\ref{c}, along the
line followed in [D].  Setting $k=-x_{1}-x_2$ in Eq.~\ref{d8c}, we get
\begin{equation}
  {\cal V}[|x_{1},x_{2}\rangle;\;x_{1}\rightarrow -x_{2},\;
   x_{2}\rightarrow -x_{1}] = -x_{1}-x_{2}+{\cal V}[|x_{1},x_{2}\rangle;\;x_{1}
   \rightarrow x{_1},\;x_{2}\rightarrow x_{2}]. \label{f}\end{equation}   
In this notation, the ``zero-sum rule'' (Eq.\ 9 of [D]) reads
\begin{equation}
  {\cal V}[|x_{1},x_{2}\rangle;\;x_{1}\rightarrow u{_1},\;
   x_{2}\rightarrow u_{2}] + {\cal V}[|x_{1},x_{2}\rangle;\;
   x_{1}\rightarrow -u{_1},\;x_{2}\rightarrow -u_{2}]=0
   \label{zs}, \end{equation}
and if we use this (with $u_{1}=-x_{2}$ and $u_{2}=-x_{1}$) in Eq.~\ref{f},
we get
\begin{equation}
   -{\cal V}[|x_{1},x_{2}\rangle;\;x_{1}\rightarrow x{_2},\;
   x_{2}\rightarrow x_{1}] = -x_{1}-x_{2} +{\cal V}
   [|x_{1},x_{2}\rangle;\;x_{1}\rightarrow x{_1},\; x_{2}\rightarrow x_{2}].
   \label{d10}\end{equation}
This is our version of Eq.\ 10 of [D].  However, we cannot now go from this
equation to Eq.~\ref{c} (i.e., to Eq.\ 11 of [D]), since the values of
the two games which appear in Eq.~\ref{d10} have not been shown to be equal
(Of course, they {\em are} equal under the condition that the two outcomes
$x_1$ and $x_2$ are equally probable, but that is precisely the condition
that [D] does {\em not} want to assume).  In fact, in the example given
by Eq.~\ref{a1} with $x_{1}=1$ and $x_{2}=0$, the outcome is surely $x_1$, 
and so
\begin{eqnarray}
   {\cal V}
   [|x_{1},x_{2}\rangle;\;x_{1}\rightarrow x{_1},\; x_{2}\rightarrow x_{2}] &
   = & x_{1}=1 \nonumber \\
   {\cal V}
   [|x_{1},x_{2}\rangle;\;x_{1}\rightarrow x{_2},\; x_{2}\rightarrow x_{1}] &
   = & x_{2}=0;  \end{eqnarray}
Eq.~\ref{d10} is satisfied (both sides equal 0), but Eq.~\ref{c} is not.

We are certainly not suggesting that Eq.~\ref{c} and the more general
Eq.~\ref{a} (i.e., Eqs. 11 and 4 of [D]) are not correct; they do follow,
after all, from the usual quantum probability rules.
Furthermore, we are not claiming to have shown that these equations are
not consequences of just the assumptions (the non-probabilistic parts of 
quantum theory and of decision theory) that were made in [D].  We do see,
however, that the arguments given in [D] are not sufficient to establish
the validity of these equations.

\vspace{1cm}
Acknowledgement: I would  like to acknowledge the hospitality of the
Lawrence Berkeley National Laboratory, where this work was done.

\end{document}